\documentclass[preprint,12pt]{elsarticle}

 \usepackage{graphicx}

\usepackage{amssymb}

\usepackage{multirow}
\usepackage{lineno}

\usepackage{color}
\usepackage{soul}

\newcommand{\Jpsi}{$J/\psi$ }
\newcommand{\pT}{$p_{_T}$ }
\newcommand{\xT}{$x_T$ }
\newcommand{\sNN}{$\sqrt{s_{_\mathrm{NN}}}$ }
\newcommand{\s}{$\sqrt{s}$ }
\newcommand{\pp}{$p$+$p$ }
\newcommand{\cucu}{Cu+Cu }
\newcommand{\auau}{Au+Au }
\newcommand{\raa}{$R_{\mathrm{AA}}$ }

\journal{Physics Letters B}

\begin{document}

\begin{frontmatter}


\title{\Jpsi production at high transverse momenta in \pp and \auau collisions at \sNN = 200~GeV}

\author{
{\Large The STAR Collaboration}
\bigskip \newline
L.~Adamczyk$^{1}$,
G.~Agakishiev$^{21}$,
M.~M.~Aggarwal$^{34}$,
Z.~Ahammed$^{53}$,
A.~V.~Alakhverdyants$^{21}$,
I.~Alekseev$^{19}$,
J.~Alford$^{22}$,
C.~D.~Anson$^{31}$,
D.~Arkhipkin$^{4}$,
E.~Aschenauer$^{4}$
% ...
}

\address{$^{1}$AGH University of Science and Technology, Cracow, Poland}
% ...
\address{$^{4}$Brookhaven National Laboratory, Upton, New York 11973, USA}
% ...
\address{$^{19}$Alikhanov Institute for Theoretical and Experimental Physics, Moscow, Russia}
% ...
\address{$^{21}$Joint Institute for Nuclear Research, Dubna, 141 980, Russia}
\address{$^{22}$Kent State University, Kent, Ohio 44242, USA}
% ...
\address{$^{31}$Ohio State University, Columbus, Ohio 43210, USA}
% ...
\address{$^{34}$Panjab University, Chandigarh 160014, India}
% ...
\address{$^{53}$Variable Energy Cyclotron Centre, Kolkata 700064, India}
% ...

%
%
%

\begin{abstract}
We report \Jpsi spectra for transverse momenta $p_{_T}>5~\textrm{GeV}/c$ at mid-rapidity
in \pp and \auau collisions at \sNN = 200~GeV.
The inclusive $J/\psi$ spectrum and the extracted $B$-hadron feed-down are compared to models
incorporating different production mechanisms.
We observe significant suppression of the \Jpsi yields for $p_{_T}\!>\!5~\textrm{GeV}/c$ in 0-30\% central \auau collisions relative to the \pp yield scaled by the number of binary nucleon-nucleon collisions in \auau collisions. In 30-60\% mid-central collisions, no such suppression is observed. The level of suppression is consistently less than that of high-\pT $\pi^{\pm}$ and low-\pT $J/\psi$ at RHIC and high-\pT \Jpsi at the LHC.
\end{abstract}


\end{frontmatter}

\section{Introduction}
\label{}
Ultrarelativistic heavy-ion collisions provide a unique
environment to study strongly interacting matter at high
temperature and energy density where a transition from the hadronic phase of matter to a new
partonic phase, the Quark-Gluon Plasma (QGP), takes place.
Measurements of the in-medium dissociation probability of the different quarkonium states are
expected to provide an estimate of the initial temperature of the system~\cite{colorscreen,sequentialSuppression1,
Mocsy:2007jz}. The $J/\psi$ is the lightest and most abundantly produced quarkonium state accessible in experiment. However, significant decay contributions ($\approx\! 40\%$~\cite{sequentialSuppression1}) from excited $c\bar{c}$ states, such as the $\chi_c$ and $\psi(2S)$, and from $B$ mesons could complicate the suppression picture suggested by dissociation models~\cite{QWG_YellowReport2010}.
In addition, other contributions absent in \pp collisions are likely to have a
significant impact on the observed \Jpsi yields in relativistic
heavy-ion collisions at CERN-SPS~\cite{Abreu:2000xe},
BNL-RHIC~\cite{PHENIX_Jpsi_AuAu} and CERN-LHC~\cite{ATLAS_PbPb_Jpsi_2011,CMS_Jpsi_PbPb_2012,ALICE_Jpsi_PbPb_2012}. These contributions include cold nuclear matter (CNM) effects such as initial state parton scattering, nuclear
shadowing and nuclear absorption, the combined contribution of finite
\Jpsi formation time and the finite space-time extent of the hot and dense volume where the dissociation can occur, and recombination
of unassociated $c$ and $\bar{c}$ in the medium~\cite{CSM_RAA1}.
Some of these processes are expected to decrease with increasing \Jpsi $p_{_T}$~\cite{Zhuang2009,XingboRalf2010}. It is therefore anticipated that \Jpsi measurements at high-\pT provide an important tool to decouple some of the mechanisms mentioned above and provide a cleaner way to extract the contribution from color-screening effects~\cite{Zhuang2009,XingboRalf2010,adscft,Sharma:2012dy}. Our previous \Jpsi measurements are consistent with no suppression for $p_{_T}\!>\!5~\textrm{GeV}/c$ in \cucu collisions at \sNN= 200~GeV, to within the limited precision of the data~\cite{starHighPtJpsiPaper}. The small system size created in \cucu collisions may result in high-\pT \Jpsi  formation outside the medium. Precise measurements in \auau collisions are thus crucial for a systematic study of \Jpsi production in the hot and dense medium.

The interpretation of medium-induced \Jpsi modification requires a good understanding of
its production mechanisms in \pp collisions, which include direct production via gluon fusion, parton fragmentation, and feed-down from higher charmonium states and $B$-hadron
decays~\cite{QWG_YellowReport2010}. The initial hard interactions required to create
the charm quark pairs can be well calculated in perturbative QCD (pQCD). However,
the subsequent soft processes required to form the \Jpsi hadron and the \Jpsi formation time are theoretically not well understood~\cite{QWG_YellowReport2010}.
No model at present fully explains the
\Jpsi observations in elementary collisions~\cite{QWG_YellowReport2010}.
The \Jpsi spectrum in the intermediate and high-\pT range, together with the angular correlations of a high-\pT \Jpsi and associated charged hadrons, may provide additional insights in the underlying production mechanisms.


In this letter, we report a measurement of \Jpsi production for $2\!<\!p_{_T}\!<\!14~\textrm{GeV}/c$ in \pp and \auau
collisions at \sNN = 200~GeV with the STAR detector~\cite{STAR_Detector_Overview}. The \Jpsi \pT spectra at mid-rapidity ($|y|\!<\!1$) in \pp and \auau collisions are
presented. The nuclear modification factor, $R_{AA}$ -- the ratio of
the yield in \auau to that in \pp collisions scaled by the number of underlying binary nucleon-nucleon collisions ($N_{bin}$ --
is calculated and compared to theoretical calculations.

\section{Data analysis}
\label{}
The \auau data used for this analysis were recorded in 2010, and the \pp data in 2009.
The minimum bias (MB) trigger was defined
to be a coincidence of the two Vertex Position Detectors~\cite{STAR_VPD_TOFp}. Online trigger conditions that utilized a MB trigger condition and two thresholds for the energy deposited in any single Barrel Electromagnetic
Calorimeter (BEMC)~\cite{STAR_BEMC} tower, with a size of
$\Delta\eta\times\Delta\phi=0.05\times0.05$, were used to maximize the sampled luminosity.
To increase the trigger efficiency, the \pp data with high
BEMC threshold were recorded without a MB requirement. The \pp data with low BEMC threshold were pre-scaled to keep the data rate manageable. The integrated luminosities of the data samples used for this analysis are 23.1~pb$^{-1}$ and 1.8~pb$^{-1}$ with a transverse energy threshold of $E_{T}\!>\!6.0$ and $2.6$~GeV, respectively, in \pp collisions and 1.4~nb$^{-1}$ with $E_{T}\!>\!4.3$~GeV in \auau collisions.
In the \auau data, the collision centrality is determined by the
distribution of charged-particle multiplicity within the pseudorapidity range $|\eta|\!<\!0.5$~\cite{starPIDLong}.

\begin{figure}[tbp]
\centering \includegraphics[width=1.0\columnwidth]{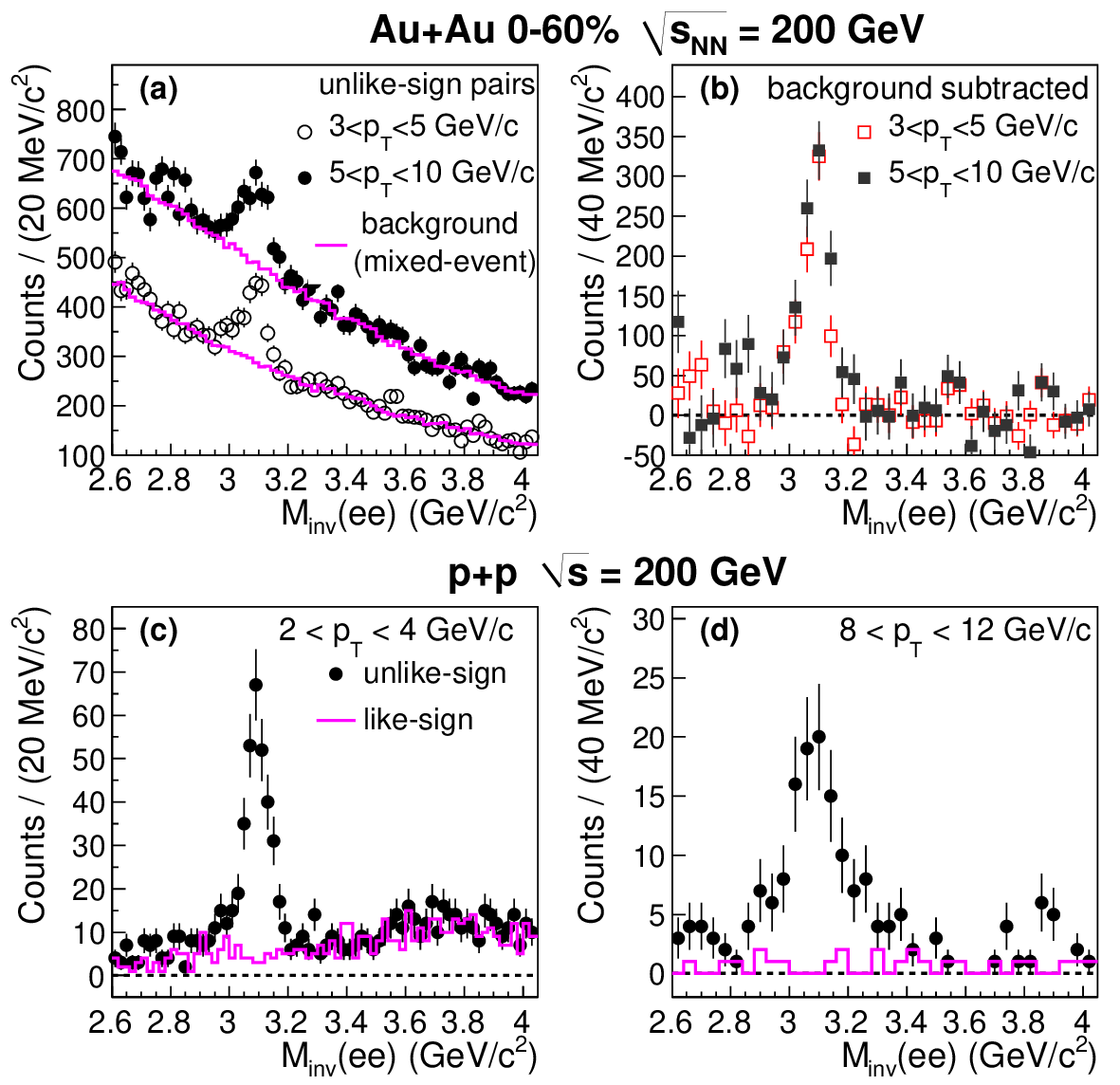}
\caption{(Color online.) (a) The unlike-sign $e^{+}e^{-}$ invariant mass distribution from same-event pairs (filled circles) and mixed-event pairs (solid curve) in \auau collisions at \sNN=200 GeV. (b) The background subtracted $e^{+}e^{-}$ invariant mass distribution in \auau collisions at \sNN=200 GeV. (c, d) The unlike-sign (filled circles) and like-sign (solid curve) $ee$ invariant mass distribution from same-event pairs in \pp collisions at \s=200 GeV. For both systems the data are shown for two intervals in \Jpsi $p_{_T}$.
} \label{fig1}
\end{figure}

\begin{figure}[tbp]
\centering \includegraphics[width=1.0\columnwidth]{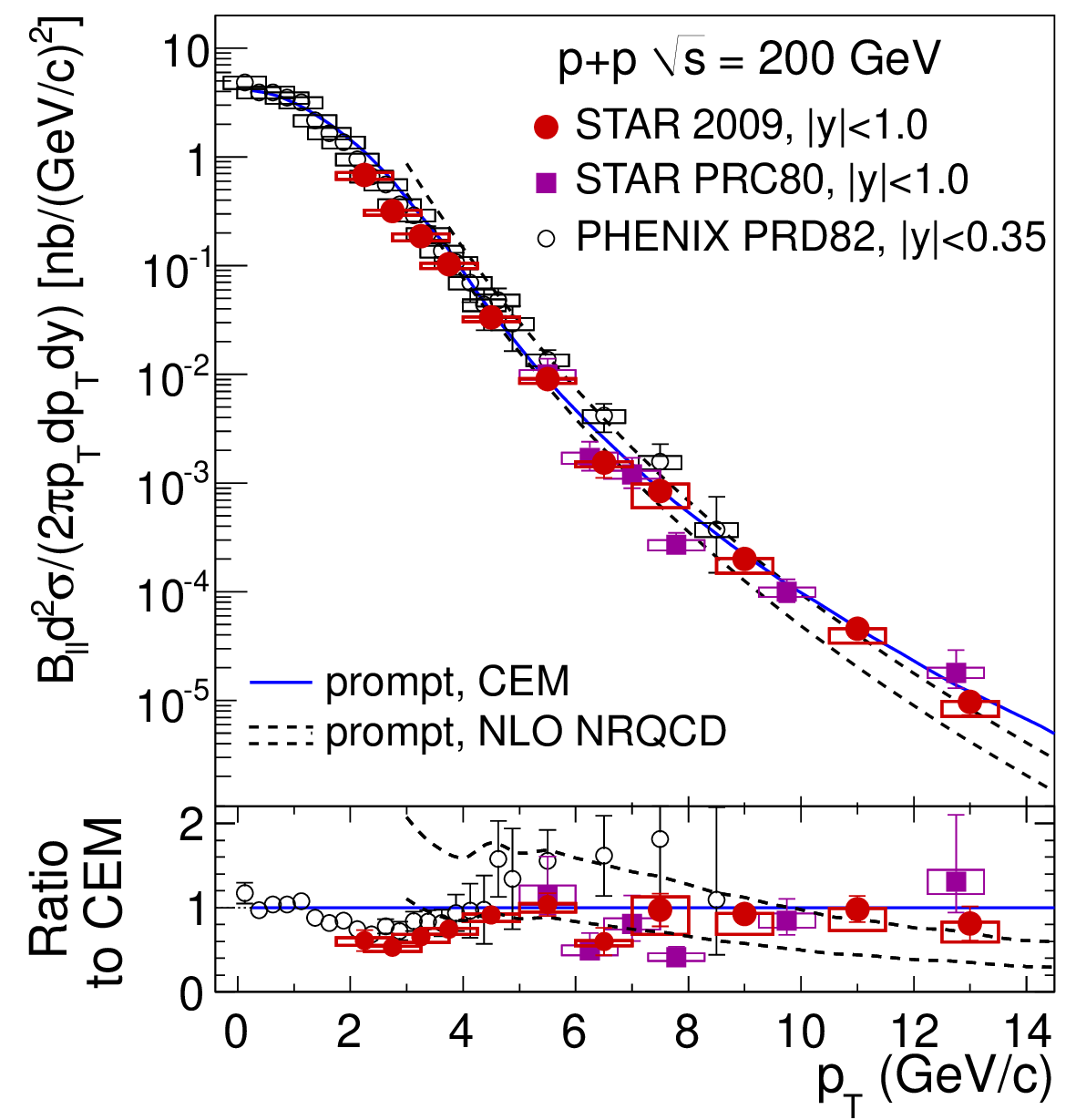}
\caption{(Color online.) The invariant \Jpsi cross section versus \pT in \pp
collisions at \s = 200~GeV. The vertical bars and boxes depict the statistical and systematic uncertainties, respectively. Also shown are results
published by STAR~\cite{starHighPtJpsiPaper} and PHENIX~\cite{Adare:2009js}.
The curves show theoretical
calculations described in the text.
} \label{fig2}
\end{figure}

\begin{figure}[tbp]
\centering \includegraphics[width=1.0\columnwidth]{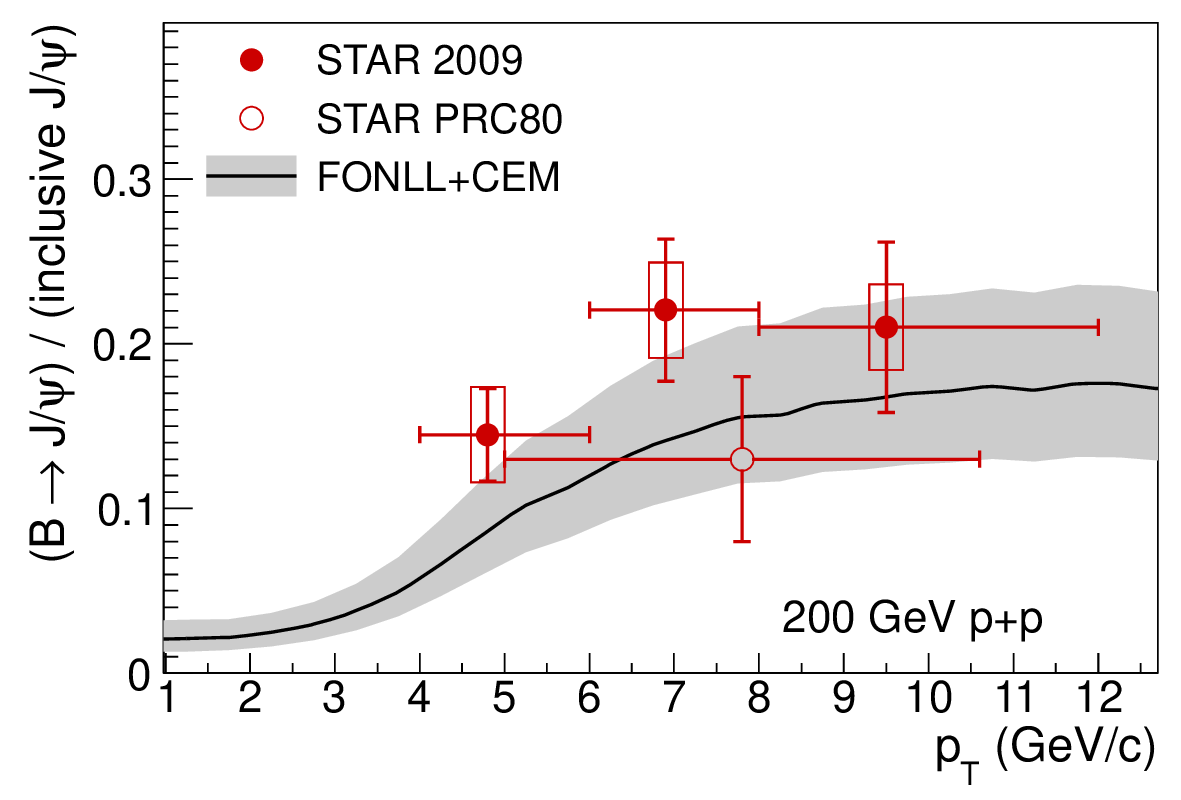}
\caption{(Color online.) The fraction of $B\rightarrow
J/\psi$ over the inclusive \Jpsi yield in \pp collisions. The FONLL+CEM model calculation is also shown.
} \label{fig3}
\end{figure}

In this analysis, \Jpsi$\rightarrow{e^+e^-}$ decays were reconstructed using the STAR Time Projection
Chamber (TPC)~\cite{STAR_TPC} and the BEMC~\cite{STAR_BEMC} with full azimuthal
coverage over the pseudorapidity range $|\eta|\!<\!1$~\cite{starHighPtJpsiPaper,zeboThesis}.
Electron identification (eID) for the BEMC triggered tracks
was achieved by measuring the ionization energy loss ($dE/dx$) and
track momentum from the TPC, as well as the energy deposition in the
BEMC. In addition, the shower profile in the barrel shower maximum detector~\cite{STAR_BEMC} was used
in \auau collisions to further suppress hadron contamination.
At moderate \pT ($1\lesssim p_{_T}\lesssim3~\textrm{GeV}/c$), TPC $dE/dx$ provides eID with reasonable efficiency and purity. At low \pT ($0.2\!<p_{_T}\lesssim3~\textrm{GeV}/c$), the eID significantly benefits from a
recently installed large area time-of-flight (TOF) detector covering $|\eta|\!<\!0.9$~\cite{STARTOF1,starTOFelectron}. The complete TOF detector was available for the 2010 \auau run, whereas 72\% was available for the \pp data collected in 2009.

The \Jpsi signal was extracted by subtracting from the unlike-sign $ee$ invariant mass spectrum the random combinatorial background that was reproduced by the like-sign spectrum in \pp
collisions and unlike-sign spectrum from mixed-events in \auau
collisions~\cite{starKstarAuAu200}. Figure \ref{fig1} shows the invariant mass distribution before and after the combinatorial background subtraction in \pp and \auau collisions at \sNN = 200 GeV. The \Jpsi raw yields were
obtained from a mass window of $2.7\!<\!M_{inv}^{ee}\!<\!3.3-3.4~\textrm{GeV}/c^2$ in \pp collisions depending on the \Jpsi $p_{_T}$, and $2.9\!<\!M_{inv}^{ee}\!<\!3.2~\textrm{GeV}/c^2$ in \auau collisions. The yields were corrected for $\approx\!10\%$ radiation losses that cause some of the decay daughters to be reconstructed with $M_{inv}^{ee}$ outside the above mass ranges. The total \Jpsi yield was $\approx\! 1100$ ($p_{_T}\!>\!2~\textrm{GeV}/c$) in \pp collisions. It was $\approx$ 1000, 600 and 300 ($p_{_T}\!>\!3~\textrm{GeV}/c$) in 0-20\%, 20-40\% and 40-60\% \auau collisions, respectively. The signal to background ratio (S/B) was $\approx\! 4$ in \pp collisions and $\approx$ 1/7 (1/2) in
0-20\% (40-60\%) \auau collisions. In the
$J/\psi$-hadron correlation analysis, different invariant mass and particle identification cuts were selected to provide a better S/B ratio. About 400 \Jpsi with $p_{_T}\!>\!4~\textrm{GeV}/c$ and $3.0\!<\!M_{inv}^{ee}\!<\!3.2$~\textrm{GeV}/$c^2$ were observed with a S/B ratio of 22/1 in \pp collisions.
\begin{figure}[tbp]
\centering
\includegraphics[width=0.95\columnwidth]{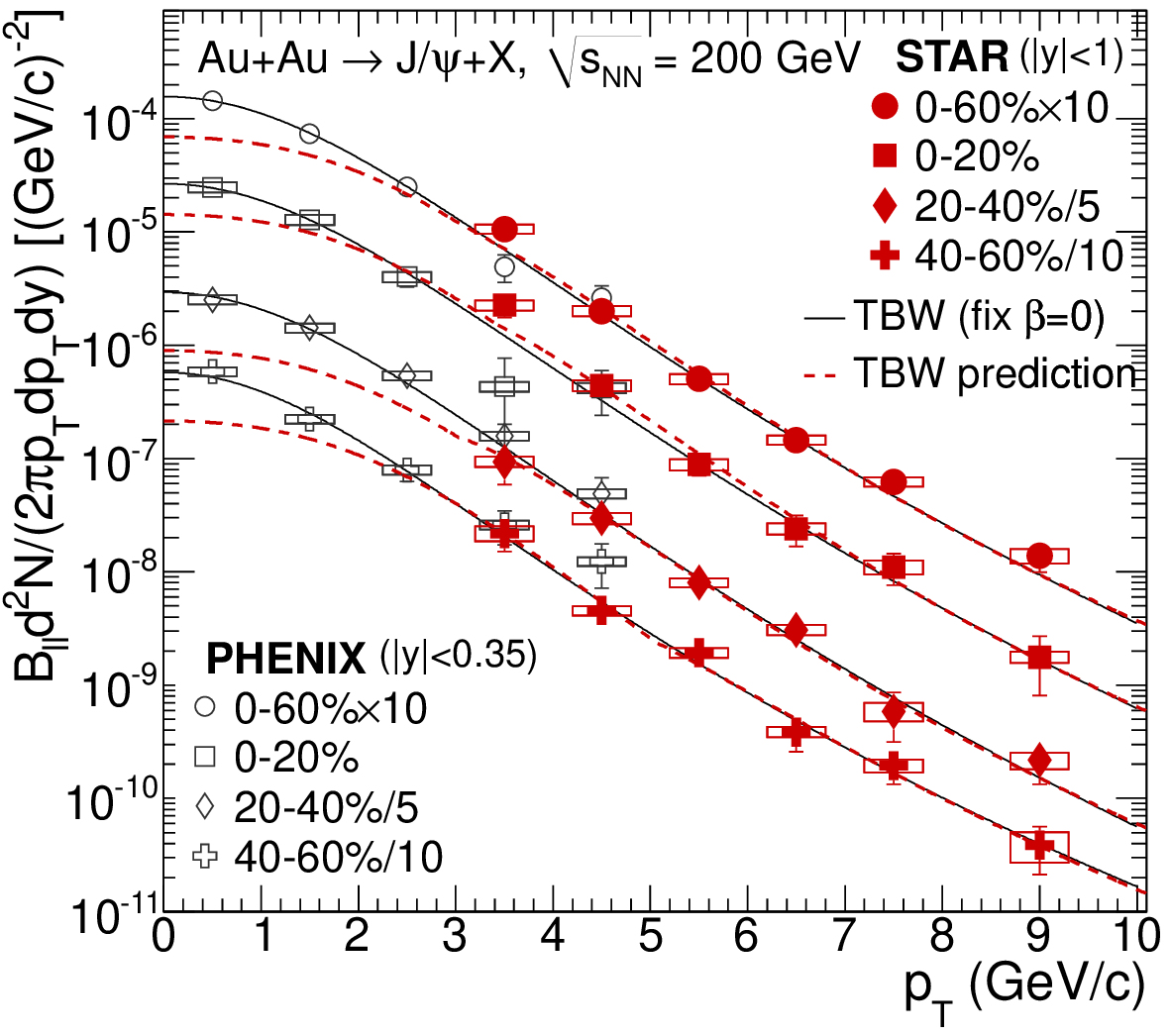}
\caption{(Color online) \Jpsi \pT distributions in \auau
collisions with different centralities at \sNN = 200~GeV.
For clarity, the data and curves have been scaled as indicated in the legends. The PHENIX results are reported in \cite{PHENIX_Jpsi_AuAu}. The curves
are model fits described in the text.}\label{fig4}
\end{figure}

Acceptance and efficiency corrections were studied using Monte Carlo (MC)
GEANT simulations~\cite{starPIDLong}.  The systematic uncertainty on this procedure is estimated by varying kinematic and eID cuts in both data analysis and MC simulations. It is uncorrelated with \pT and collisions systems. An additional systematic uncertainty of 7.5\%, obtained from two complementary simulation methods, was assigned for the \pT spectra in \pp and \auau collisions. This contribution cancels in the uncertainty on $R_{\mathrm{AA}}$. We have also varied the invariant mass window for signal counting to evaluate the systematic uncertainty on the yield extraction procedure, including the contributions from radiation losses and correlated background. This contribution is larger in \pp than in \auau collisions due to the wider mass window used in \pp collisions. Since it is correlated in \pp and \auau collisions, this systematic uncertainty on \raa is estimated by varying the mass window in the same way in \pp and \auau collisions. The normalization uncertainty for the cross section in \pp collisions is 8.1\%~\cite{ppUncertainty}. The normalization uncertainty for \raa has also a contribution from the uncertainty on the calculation of $N_{bin}$. The combined uncertainty varies from 8.6\% in 0-10\% central \auau collisions to 23.4\% in 40-60\% central \auau collisions. The systematic uncertainties are summarized in Tab.~\ref{tab:tab1}.

\begin{table}
\caption{Summary of assigned systematic uncertainties of \Jpsi spectra in \pp and \auau collisions, and \Jpsi \raa in \auau collisions.}\label{tab:tab1}
\begin{tabular}{lcccc}
  \hline
  Description & \pp & \auau & \raa \\
  \hline
  Kinematic and eID cuts &1-16\% & 3-28\% & 3-28\% \\
  Efficiency &   7.5\% &  7.5\% &  - \\
  Yield extraction & 8-26\% & 2\% & 4\%\\
  Normalization & 8.1\% & - & 8.6-23.4\%\\
  \hline
\end{tabular}
\end{table}
%


\section{Results and discussion}
Figure~\ref{fig2} shows the
\Jpsi invariant cross-section times the branching ratio ($B_{ll}$)~\cite{PDG2010}
at mid-rapidity ($|y|\!<\!1$) as a function of \pT for \pp collisions at \s = 200~GeV. The new results are consistent with those previously
published by STAR~\cite{starHighPtJpsiPaper} and PHENIX~\cite{Adare:2009js}. The dashed curves depict the uncertainty band of next-to-leading order (NLO) theoretical Non-Relativistic QCD (NRQCD) calculations from
color-octet (CO) and color-singlet (CS) transitions~\cite{NRQCD_PKU1012} for prompt \Jpsi production in \pp collisions. The CS+CO calculations match the \pT spectra for $p_{_T}\!>\!$ 4~GeV/$c$ to within the uncertainties. The solid curve
shows the calculation from the color evaporation model (CEM) for
prompt $J/\psi$~\cite{Frawley:2008kk}. It describes the \pT spectra reasonably well at low and high $p_{_T}$, but overpredicts the data by a factor of 2 at \pT around 3 GeV/$c$. The bottom panel shows the ratios of the data and NRQCD calculations to the CEM calculation.
Not shown in this figure is the model based on NNLO$^{\star}$ CS~\cite{Artoisenet:2008fc}, which predicts a too steep \pT
dependence, as discussed in~\cite{starHighPtJpsiPaper}.

\begin{figure}[tbp]
\centering \includegraphics[width=1.\columnwidth]{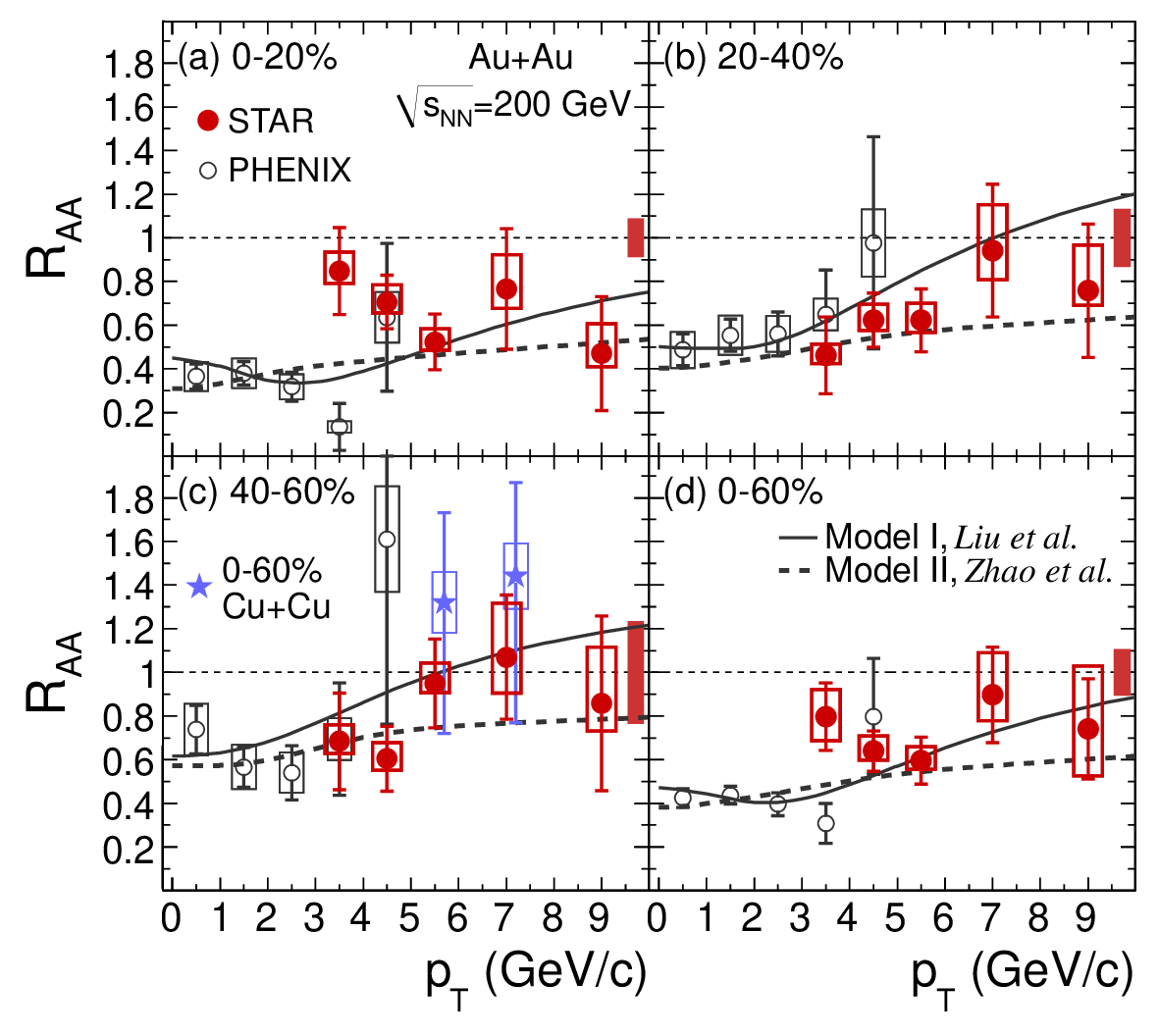}
\caption{(Color online.) \Jpsi \raa versus \pT for several centrality bins for \auau
collisions at \sNN = 200~GeV. The statistical (systematic) uncertainties are shown
with vertical bars (open boxes). The filled boxes about unity on the right show the size of the normalization uncertainty. PHENIX low-\pT \Jpsi results~\cite{PHENIX_Jpsi_AuAu} and STAR high-\pT results in \cucu collisions~\cite{starHighPtJpsiPaper} are shown for comparison.
The curves are from the predictions by Model I (Liu et al.)~\cite{Zhuang2009} and Model II (Zhao et al.)~\cite{XingboRalf2010}.} \label{fig5}
\end{figure}

The relative contribution of $B$-hadron feed-down to the inclusive \Jpsi yield was obtained by fitting the azimuthal angular correlation between high-\pT \Jpsi and charged hadrons with simulated correlation functions for prompt \Jpsi and \Jpsi from $B$-hadron feed-down from PYTHIA~\cite{Pythia6,Pythia8}. Details of this procedure are described in \cite{starHighPtJpsiPaper,zeboThesis}. The separation of the correlation functions for the two above contribution sources increases as a function of $p_{_T}$. We note that this method is data-driven, although it relies also on the validity of PYTHIA's modeling of the near-side associated hadron distri\-butions. The $B$-hadron feed-down contribution is found to be within 10-25\% in the range $4\!<\!p_{_T}\!<\!12~\textrm{GeV}/c$ as shown in Fig.~\ref{fig3}. Within err\-ors our data are consistent with the Fixed Order plus Next-to-Leading Logarithms (FONLL) plus CEM prediction~\cite{Bedjidian:2004gd,FONLL} indicated by the curve and uncertainty band. More precise measurements using displaced vertex techniques~\cite{STAR_HFT2} similar to those employed by CDF in $p+\bar{p}$ collisions
at \s= 1.96~TeV and by ATLAS and CMS in \pp collisions at \s= 7~TeV~\cite{JpsiSpectra_CDFII,ATLASJpsi,CMSJpsi} are needed to quantify the anticipated energy dependence~\cite{Bedjidian:2004gd,FONLL}.

The measured \Jpsi \pT spectra in \auau collisions for different centralities are shown in Fig.~\ref{fig4}. The shape of the \Jpsi $p_{_T}$-distribution depends not only on the production mechanism but in heavy ion collisions also on the level of charm quark thermalization. \Jpsi produced from direct pQCD processes have no initial collective motion and may acquire radial flow through interaction with the hot and dense medium. \Jpsi produced from charm quark recombination should inherit the flow of charm quarks and exhibit an enhancement of yield at low $p_{_T}$. Thus the study of \Jpsi \pT spectra with a thermal model may provide insight in its production mechanism and thermalization. The solid curves
depict fits based on the Tsallis statistics Blast-wave (TBW)
model to the combined STAR and PHENIX data with the
radial flow velocity $\beta$ fixed to zero~\cite{Tsallis09}. The fits
reproduce the data reasonably well. Under the assumption that the \Jpsi flows like light hadrons~\cite{Tsallis09, Tsallis11},
the TBW calculations shown as dashed curves underpredict the yields at low $p_{_T}$.  This could be due to a
small (or zero) radial flow or a significant contribution from
charm quark recombination that would enhance the yield at low $p_{_T}$, or both.

\begin{figure}[tbp]
\centering
\includegraphics[width=1.\columnwidth]{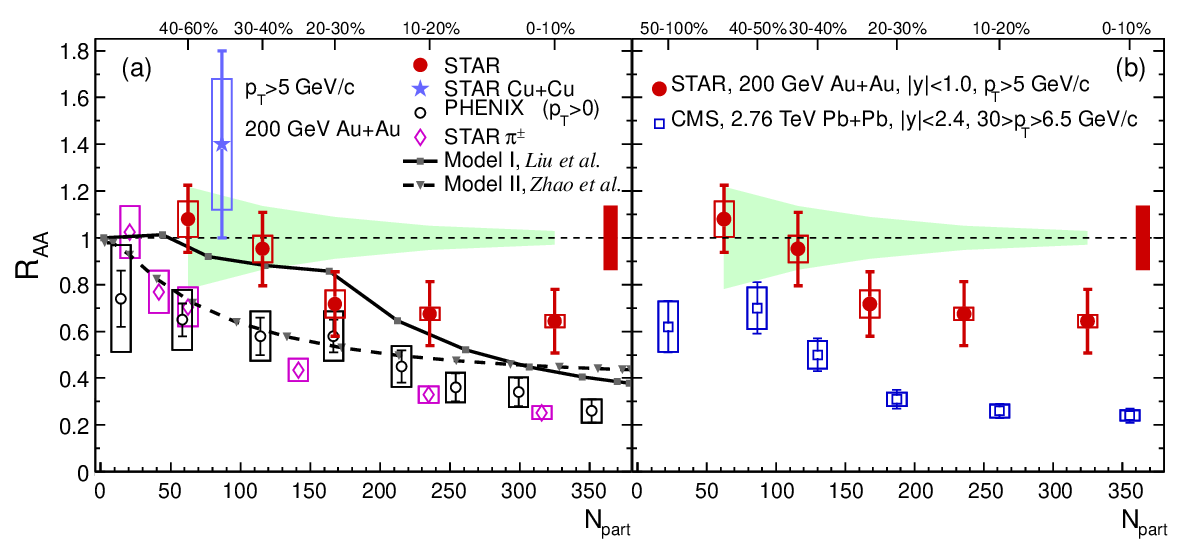}
\caption{(Color online) \raa versus $N_{part}$ for high-\pT $J/\psi$ in comparison with a) low-\pT \Jpsi data from PHENIX~\cite{PHENIX_Jpsi_AuAu}, and high-\pT $\pi^{\pm}$ from STAR \cite{YichunPIDPRL,LijuanPIDPRL}, and b) high-\pT \Jpsi from CMS~\cite{CMS_Jpsi_PbPb_2012}. The statistical (systematic) uncertainties are shown in vertical bars (boxes). The shaded green band about unity shows the systematic uncertainties from $N_{bin}$ and the box about unity on the right shows the \raa normalization uncertainty from the statistical and global systematic uncertainties of the \pp reference data. The percentage in the top horizontal axis of (a) and (b) refer to the collision centrality for high-\pT \Jpsi data at the RHIC and LHC, respectively.
}\label{fig6}
\end{figure}

Figure~\ref{fig5} shows the \Jpsi \raa versus \pT for diff\-er\-ent
centrality bins. The STAR
\cucu \Jpsi results~\cite{starHighPtJpsiPaper} are also shown.
For $p_{_T}\!>\!5~\textrm{GeV}/c$, \Jpsi \raa in the 40-60\%
centrality bin is consistent with unity and with our previous measurement in \cucu
collisions with a similar average number of participants ($N_{part}$). The curves show two theoretical calculations~\cite{Zhuang2009,XingboRalf2010} describing the data reasonably well. These calculations include contributions from prompt production and statistical charm quark regeneration. The suppression of the
prompt \Jpsi component in the model calculations is mainly due
to the color-screening effect. The model from Zhao et al. (Model II)~\cite{XingboRalf2010} also includes the \Jpsi formation-time effect and the $B$-hadron feed-down contribution.

For $p_{_T}\!>\!5~\textrm{GeV}/c$, \Jpsi production follows the scaling of the cross section at mid-rapidity, $\frac{d^2\sigma}{2\pi p_{_T} dp_{_T} dy}=g(x_T)/(\sqrt{s})^{n}$, where $x_{_T}=2p_{_T}/\sqrt{s}$ and $g(x_{_T})$ is a universal function of $x_{_T}$~\cite{xT_scaling_history1,scalingpi,LijuanPIDPLB,xT_scaling_PHENIX} observed for a wide range of collision energies and confirmed by \-STAR data on \pp collisions at \s = 200~GeV~\cite{starHighPtJpsiPaper}. It has been argued that the breaking of \xT scaling at low \pT ($p_{_T}\!\approx\!2~\textrm{GeV}/c$) for hadrons (pions and protons) \cite{starHighPtJpsiPaper,scalingpi,LijuanPIDPLB,xT_scaling_PHENIX} indicates a possible transition from soft to hard processes and thus defines the range where pQCD is applicable. Similarly, the observed \xT scaling behavior of the \Jpsi suggests a transition from soft to hard process at around $p_{_T}\!\approx\!5~\textrm{GeV}/c$. In the high \pT region, the CNM and $c\bar{c}$ recombination effects
are expected to be negligible in heavy-ion collisions~\cite{Zhuang2009,XingboRalf2010}. By consequence, \Jpsi suppression in this \pT region should provide a cleaner probe of the hot medium suppression effects. Furthermore, a color-screening model based on AdS/CFT predicts that the dissociation temperature of direct \Jpsi in QGP decreases to 1.5 $T_c$, which is believed to be reached in RHIC collisions, at $p_{_T}\!>\!5~\textrm{GeV}/c$~\cite{adscft}.

We present high-\pT ($p_{_T}\!>\!5~ \textrm{GeV}/c$) \raa as a function of $N_{part}$ in Fig.~\ref{fig6}. No significant suppression of high-\pT \Jpsi production is observed in mid-central to peripheral collisions (30-60\%, $N_{part}\lesssim 140$). However, in central collisions
(0-30\%, $N_{part}\gtrsim 140$), high-\pT \Jpsi are significantly suppressed. The \raa of low-\pT ($0\!<\!p_{_T}\!<\!5$~GeV/$c$) \Jpsi measured by PHENIX~\cite{PHENIX_Jpsi_AuAu} and high-\pT ($p_{_T}\!>\!5$~GeV/$c$)
charged pions measured by STAR~\cite{YichunPIDPRL,LijuanPIDPRL} are shown in Fig. \ref{fig6}(a) for comparison. The high-\pT \Jpsi \raa is
systematically higher. Note, that the $B$-hadron feed-down contribution is not subtracted and that part of the suppression could for instance reflect $b$-quark energy loss. Based on our measurement, \raa for prompt \Jpsi with $p_{_T}\!>\!5~\textrm{GeV}/c$ in the most central collisions will be $0.80\pm0.17$ if $B$-hadron $R_{\mathrm{AA}}\!=\!0$ and $0.55\pm0.17$ if $B$-hadron $R_{\mathrm{AA}}\!=\!1$. The predictions of high-\pT \Jpsi \raa from Model I (Liu et al.)~\cite{Liu:2009gx} and Model II (Zhao et al.)~\cite{XingboRalf2010} are shown as solid and dashed curves, respectively. Model I describes our data reasonably well. Model II underpredicts \Jpsi \raa ($p$-value=0.0018 with all of the uncertainties taken into account). Fig.~\ref{fig6}(b) shows the comparison of high-\pT \Jpsi \raa versus $N_{part}$ from STAR at RHIC and CMS at the LHC~\cite{CMS_Jpsi_PbPb_2012}. The STAR results are higher for all centralities.

The high-\pT \Jpsi \raa versus centrality is different from that of high-\pT pions. This is expected from differences in their production. Dissociation is considered to be the dominant mechanism that determines the \raa in the case of \Jpsi production, and induced gluon radiation in the case of pion production. For $p_{_T}\!>\!5~\textrm{GeV}/c$, the
recombination and initial parton scattering effects are expected to be negligible~\cite{Zhuang2009,XingboRalf2010}. The observed \Jpsi
\raa dependence on system size in this \pT range might be due to the
interplay of formation time, color screening and parton distribution functions in heavy
nuclei~\cite{CSM_RAA1}. The model calculations of \cite{XingboRalf2010} include formation time effects and CNM effects, and predict that \Jpsi \raa is close to unity for $p_{_T}>5$ GeV$/c$. Therefore significant suppression observed in central 0-30\% \auau collisions points to the color screening features. Empirically, the \Jpsi is the only measured hadron that exhibits significant suppression in \raa and has neither observable elliptic flow~\cite{Jpsi_v2_STAR} nor model-extracted radial flow~\cite{Tsallis09} at RHIC. The strong suppression of \Jpsi at high \pT and the sequential $\Upsilon$ suppression observed at LHC by the CMS Collaboration~\cite{CMS_Jpsi_PbPb_2012,CMS_Upsilon_doubleRatio_PbPb_2011,CMS_Upsilon_QM2012} are consistent with this interpretation. Comparison of low-\pT \Jpsi yields at RHIC and LHC~\cite{Jpsi_ALICE_Pt_2012} and study of the \Jpsi azimuthal anisotropy~\cite{Jpsi_v2_STAR,Jpsi_v2_Alice_Yang:2012fw} could quantitatively further constrain the model interpretation.

\section{Summary}
This letter reports measurements of \Jpsi production in \sNN
= 200~GeV \pp and \auau collisions for $p_{_T}\!>\!2-3$~GeV/$c$
at RHIC. The \pT spectrum in \pp collisions is compared to
various theoretical cal\-culations. Currently, only the CEM model and NLO CS+CO calculation describe our data. Based on the measurement of azimuthal correlations between high-\pT \Jpsi and charged hadrons we estimate the fraction of \Jpsi from $B$-hadron decay to be 10-25\% in the \pT range of 4-12~GeV/$c$ in \pp
collisions. We report the first measurement of high-\pT \Jpsi suppression in \auau collisions at RHIC. The nuclear modification factor \raa in \auau increases from low to high $p_{_T}$. For $p_{_T}\!>\!5$~\textrm{GeV}/$c$, $J/\psi$ \raa is consistent with no suppression from mid-central to peripheral collisions (30-60\% centrality), and significantly smaller than unity in the most central \auau collisions. The results on \raa versus \pT and $N_{part}$ provide new insight in the study of color screening features for charmonium.

\section*{Acknowledgments}

We thank the RHIC Operations Group and RCF at BNL, the NERSC Center at LBNL and the Open Science Grid consortium for providing resources and support. This work was supported in part by the Offices of NP and HEP within the U.S. DOE Office of Science, the U.S. NSF, the Sloan Foundation, the DFG cluster of excellence `Origin and Structure of the Universe' of Germany, CNRS/IN2P3, FAPESP CNPq of Brazil, Ministry of Ed. and Sci. of the Russian Federation, NNSFC, CAS, MoST, and MoE of China, GA and MSMT of the Czech Republic, FOM and NWO of the Netherlands, DAE, DST, and CSIR of India, Polish Ministry of Sci. and Higher Ed., Korea National Research Foundation, Ministry of Sci., Ed. and Sports of the Rep. of Croatia, and RosAtom of Russia.




\bibliographystyle{elsarticle-num}
\bibliography{highPtJpsi_pp_AuAu_STAR_v3}







\end{document}